\documentclass[10pt]{article}
\usepackage[top=2cm,bottom=3cm,left=2.5cm,right=2.5cm,marginparwidth=1.75cm]{geometry}
\usepackage{graphicx} 
\usepackage{amssymb}
\usepackage{amsmath,bm}
\usepackage{multicol}
\usepackage{appendix}

\newcommand\Ha{\mbox{\rm{Ha}}}  
\newcommand\Nu{\mbox{\rm{Nu}}}  
\newcommand\Ra{\mbox{\rm{Ra}}}  
\newcommand\Pran{\mbox{\rm{Pr}}} 
\newcommand\Rey{\mbox{\rm{Re}}}  

\title{\Large{\textbf{Unifying heat transport model for the transition between buoyancy-dominated and Lorentz-force-dominated regimes in quasistatic magnetoconvection} }}
\author{Andrei Teimurazov\textsuperscript{1}, Matthew McCormack\textsuperscript{2,}\footnote{A. Teimurazov and M. McCormack contributed equally.}, Moritz Linkmann\textsuperscript{2,}\footnote{moritz.linkmann@ed.ac.uk}, and Olga Shishkina\textsuperscript{1,}\footnote{olga.shishkina@ds.mpg.de}}
\date{\small{\textsuperscript{1}{Max Planck Institute for Dynamics and Self-Organization, 37077 Göttingen, Germany}\newline {\textsuperscript{2}School of Mathematics and Maxwell Institute for Mathematical Sciences, University of Edinburgh, UK}}\newline \newline August 3, 2023}

\begin{document}

\maketitle

\vspace{-20pt}
\begin{abstract}
In magnetoconvection, the flow of electromagnetically conductive fluid is driven by a combination of buoyancy forces, which create the fluid motion due to thermal expansion and contraction, and Lorentz forces, which distort the convective flow structure in the presence of a magnetic field.
The differences in the global flow structures in the buoyancy-dominated and Lorentz-force-dominated regimes lead to different heat transport properties in these regimes, reflected in distinct dimensionless scaling relations of the global heat flux (Nusselt number $\Nu$) versus the strength of buoyancy (Rayleigh number $\Ra$) and electromagnetic forces (Hartmann number $\Ha$).
Here, we propose a theoretical model for the transition between these two regimes for the case of a quasistatic vertical magnetic field applied to a convective fluid layer confined between two isothermal, a lower warmer and an upper colder, horizontal surfaces.
The model suggests that the scaling exponents $\gamma$ in the buoyancy-dominated regime, $\Nu\sim\Ra^\gamma$, and $\xi$ in the Lorentz-force-dominated regime, $\Nu\sim(\Ha^{-2}\Ra)^\xi$, are related as $\xi=\gamma/(1-2\gamma)$, and the onset of the transition scales with $\Ha^{-1/\gamma}\Ra$.
These theoretical results are supported by our Direct Numerical Simulations for $10\leq \Ha\leq2000$, Prandtl number $\Pran=0.025$ and $\Ra$ up to $10^9$ and data from the literature.
\end{abstract}

\vspace{5pt}
\section{{Introduction}}
Magnetoconvection (MC) governs most astro- and geophysical systems and is relevant to various engineering applications \cite{Weiss2014, Davidson2016}. The former include, for instance, outer layers of stars and liquid metal planetary cores \cite{Jones2011}, examples of the latter comprise liquid-metal batteries, induction heating, casting, liquid-metal cooling for nuclear fusion reactors and semiconductor crystal growth \cite{Davidson1999}.  
MC occurs in an electrically conducting fluid that is subjected both to a magnetic field and an imposed temperature gradient.  
The buoyancy forces induce convective fluid motion due to thermal expansion and contraction, while the magnetic field affects this motion
and distorts the global flow structure through the Lorentz force, which eventually influences the heat transport in the system. 
The resulting main two control parameters, the strength of the imposed thermal driving and that of the external magnetic field, are encoded in independent dimensionless groups, the Rayleigh number $\Ra$ and Hartmann number $\Ha$, respectively.

One of the key objectives in MC research is to provide scaling relations for the heat transport through the system, represented in dimensionless form by the Nusselt number $\Nu$, as a function of $\Ra$ and $\Ha$.
However, the heat transport scaling relations also depend on the flow configuration, including the angle between the magnetic field and gravity, the geometry of the container and the boundary conditions (BCs), and on whether the buoyancy forces dominate over the Lorentz forces in the system or vice versa. 
This inherent complexity results in the need, at least in principle, to derive separate heat transport scaling relations to describe each specific flow regime itself and transitions between distinct regimes.  
The considerable difficulty of doing so in a coherent manner is exacerbated by non-universal scaling relations even within specific regimes -- the scaling relations in the buoyancy-dominated and Lorentz-force-dominated regimes themselves change with the control parameters, and transitions between the different regimes are also non-universal.

The objective of this paper is to offer a unifying heat transport model for the transition between the buoyancy-dominated and Lorentz-force-dominated regimes in quasistatic MC.  
We focus on Rayleigh--B\'enard convection (RBC) \cite{Ahlers2009} with an applied vertical magnetic field and assume that the magnetic field is constant in the entire domain, without being affected by a fluid motion or finite magnetic diffusion.  
The model uses the theoretical predictions by Grossmann and Lohse \cite{Grossmann2000, Grossmann2001, Stevens2013} for RBC without magnetic field and transfers the approach by Ecke and Shishkina \cite[\S3.3]{Ecke2023} for transitions in rotating RBC to the case of RBC with a vertical magnetic field.  
To verify the proposed model, we compare the theoretical predictions with results for liquid metal MC obtained by direct numerical simulation (DNS) carried out by us and others \cite{Liu2018, Yan2019, Akhmedagaev2020, Xu2023}, as well as experiments \cite{Cioni2000, King2015, Zuerner2020, Xu2023}. 
In addition, we carried out simulations for a different working fluid at higher Prandtl number $\Pran$ to compare our DNS data with that from Ref.~\cite{Lim2019}.  
The predictions of the proposed model agree well the experimental and DNS data.

\section{{Model for the transition between the buoyancy dominated and Lorentz force dominated regimes}}
We consider a layer of electrically conducting fluid confined between two infinitely wide and long plates, driven by a buoyancy force generated by an imposed vertical temperature difference between the top and bottom plates, and subjected to a uniform vertically orientated magnetic field.  
When the magnetic field is weak compared to the buoyancy force, we recover classical RBC scaling for the dimensionless convective heat flux $\Nu-1$, that is, the total dimensionless heat flux $\Nu$ less its conductive contribution, $\Nu - 1 \sim (\Ra/\Ra_{c,b})^\gamma \sim \Ra^\gamma$, for an exponent $\gamma$, where where $\Ra_{c,b}$ is the critical $\Ra$ for RBC bulk onset for a given container geometry. 
When the Lorentz force is strong compared to the buoyancy force, we expect a similar scaling law $\Nu - 1 \sim \left(\Ra/\Ra_{c,L}\right)^\xi$, for an exponent $\xi$. 
Here, the dependence on the critical Rayleigh number $\Ra_{c,L}$ in the Lorentz-force-dominated regime is kept due to its dependence on $\Ha$, that can be obtained from the linear stability theory \cite{Chandrasekhar1961} $\Ra_{c,L} \sim \Ha^2$, with no dependence on the Prandtl number.

\begin{figure}
\centering
\includegraphics[width = 16cm]{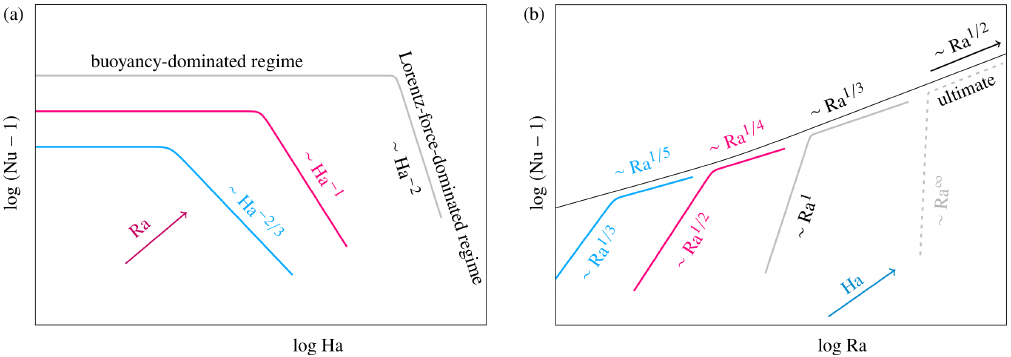}
\caption{Scalings of $\Nu-1$ versus (a) $\Ha$ and (b) $\Ra$, according to the theory.}
\label{Fig1}
\end{figure}

Although these buoyancy-dominated and Lorentz-force-dominated scaling laws appear disconnected, they are intrinsically linked under the assumption that they must overlap at some intermediate region between the two extreme regimes, where neither the influence of the Lorentz force or buoyancy force on the convective heat transport can be ignored.  
At the transition between these regimes, the two corresponding scaling laws will scale in the same way,
\begin{equation}
    \Ra^\gamma\sim\Nu-1\sim(\Ha^{-2}\Ra)^\xi, 
    \label{qqq1}
\end{equation}
To construct a relationship between the two scaling exponents $\gamma$ and $\xi$, we assume that this cross-over occurs when the thicknesses of the thermal and viscous boundary layers (BLs) scale in the same way, motivated by the observation that the viscous layer is nested within the thermal BL in the Lorentz-force-dominated regime and {\em vice versa} in the buoyancy-dominated regime.
In a domain of height $H$ over a semi-infinite plate, the average thermal BL thickness from laminar Prandtl-Blasius boundary layer theory \cite{Prandtl1905, Schlichting1979} is $\delta_\theta = H/(2\Nu) \sim \Nu^{-1} $, while the laminar viscous BL influenced by a vertical magnetic field is the Hartmann layer \cite{Hartmann1937b, Davidson2016}, $\delta_{\small \Ha} = c H/\Ha \sim \Ha^{-1} $, where $c$ is a constant.
Assuming that $\delta_\theta \sim \delta_{\small \Ha}$ at the transition implies $\Nu \sim \Ha$. 
Since this transition is seen to typically occur at high Nusselt numbers, meaning $\Nu \approx \Nu-1$, we obtain
\begin{equation}
\Ha \sim \Nu \sim \Ra^\gamma \sim \Ra^\xi \Ha^{-2\xi} \sim \Ra^{-2\xi\gamma+\xi},
\end{equation}
resulting in the following relationship between the exponents,
\begin{equation}
\xi={\gamma/(1-2\gamma)}, \quad \textrm{or}\quad 
\gamma={\xi/(1+2\xi)}.  
\label{qqq5}
\end{equation}
One can see that a larger (smaller) exponent in one regime requires a larger (smaller) exponent the other regime, and that $\gamma$ is always smaller than $1/2$.

\begin{figure}[h]
\centering
\includegraphics{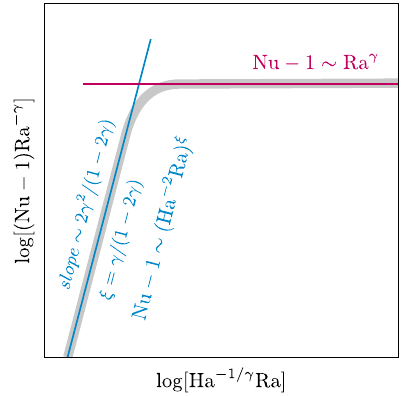}
\caption{Schematic representation of the normalized convective heat transport $\Nu-1$ displaying the transition from the Lorentz-force-dominated regime, $\Nu-1\sim(\Ha^{-2}\Ra)^\xi$, to the buoyancy-dominated regime, $\Nu-1\sim \Ra^\gamma$, according to our model. 
The scaling exponents $\xi$ and $\gamma$ follow Eq.~(\ref{qqq5}), while the transition scales with $\Ha^{-1/\gamma}\Ra$.}
\label{Fig2}
\end{figure}

In Fig.~\ref{Fig1} we present a sketch of the proposed scaling relations for $(\Nu-1)$ versus $\Ha$ (Fig.~\ref{Fig1}~a) and $\Ra$ (Fig.~\ref{Fig1}~b), according to the relations (\ref{qqq5}).  
Once $\gamma$ is known for any specific $\Pran$, the exponent $\xi$ can be calculated from Eq.~(\ref{qqq5}).
These scalings can then be used define coordinates $(\Nu-1)\Ra^{-\gamma}$ and $\Ha^{-1/\gamma}\Ra$ with respect to which the heat transport dependence for different values of $\Ha$ and $\Ra$ collapse onto a master curve, as sketched in Fig.~\ref{Fig2}.  
The transition then should take place in a Rayleigh-number range that scales as $\Ha^{1/\gamma}$. 

To close the model, a theoretical prediction for either $\gamma$ or $\xi$ must be made.  
The former is readily available through Grossmann--Lohse (GL) theory \cite{Grossmann2000, Grossmann2001, Stevens2013} for RBC without magnetic field, applicable here in the buoyancy-dominated regime.  
For any given $\Pran$ and $\Ra$-range, the theory provides accurate predictions of the value of $\gamma$, for containers of aspect ratio
$\Gamma \gtrsim 1$.  
For $\Gamma\ll 1$, the data can be rescaled according to the method suggested in \cite{Shishkina2021, Ahlers2022}, which we do not discuss here, as in the present study $\Gamma=1$.

\newpage
\section{{Status Quo -- experimental and numerical data.}}
To verify the theoretical model we compare its predictions against data obtained from experiments of liquid-metal MC \cite{Cioni2000, King2015, Zuerner2020, Xu2023} and DNS conducted by us and others \cite{Liu2018, Yan2019, Akhmedagaev2020, Xu2023}.  
However, before doing so we provide a brief overview of the data collated from the literature and produced by us to demonstrate the considerable challenges that arise when trying to draw firm conclusions on the scaling of the heat transport with $\Ha$ and $\Ra$. 

We simulate an incompressible, viscous buoyancy-driven flow of an electrically conducting fluid in the presence of an imposed magnetic field for very small magnetic Reynolds number $\Rey_m \ll 1$ and magnetic Prandtl number $\Pran_m \ll 1$ by numerically solving the MC equations within the Oberbeck--Boussinessq and quasistatic approximation,
\begin{eqnarray}
\frac{\partial \bm{u}}{\partial t} + \bm{u}\cdot {\nabla} \bm{u} + {\nabla} p &=& \sqrt{\frac{\Pran}{\Ra}} [{\nabla}^2 \bm{u} + \Ha^2(\bm{j} \times \bm{e}_B)] + T \bm{e}_z,\quad \ \ \label{eq:MHD1}\\
\frac{\partial T}{\partial t} + \bm{u}\cdot {\nabla} T &=&  \frac{1}{\sqrt{\Ra \Pran}} {\nabla}^2 T,\label{eq:MHD2}\\
{\nabla} \cdot \bm{u} &=& 0,\label{eq:MHD3}\\
\bm{j} &=& -{\nabla} \phi + \bm{u} \times \bm{e}_B, \label{eq:MHD4} \\
{\nabla}^2 \phi &=& {\nabla} \cdot (\bm{u} \times \bm{e}_B) \label{eq:MHD5}.
\end{eqnarray}
where $\bm{u}$ is the velocity, $T$ the temperature, $p$ the kinematic pressure, $\bm{j}$ the electric current density, $\phi$ the electric potential, and $\bm{e}_z$ and $\bm{e}_B$ are unit vectors that point, respectively, upward (opposite to gravity) and in the direction of the magnetic field $\bm{B} = B_0 \bm{e}_B$.
The magnetic field is aligned with the buoyancy force, $\bm{e}_B=\bm{e}_z$. 

Equations ~(\ref{eq:MHD1})--(\ref{eq:MHD5}) have been non-dimensionalized using the container height $H$, the free-fall velocity $u_{f\!f}\equiv(\alpha gH\Delta)^{1/2}$, the free-fall time $t_{f\!f}\equiv H/u_{f\!f}$, the temperature difference between the bottom and top plates, $\Delta\equiv T_+-T_-$, and the external magnetic field strength, $B_0$, as units of length, velocity, time, temperature and magnetic field strength, respectively.
The dimensionless control parameters are the Rayleigh number $\Ra$, the Prandtl number $\Pran$, and the Hartmann number $\Ha$,
\begin{eqnarray}
\Ra\equiv~\frac{\alpha g \Delta H^3}{\kappa \nu}, \qquad \Pran\equiv~\frac{\nu}{\kappa}, \qquad \Ha\equiv~B_0 H \sqrt{\frac{\sigma}{\rho \nu}} \ , \label{gov_par}    
\end{eqnarray}
where  $\sigma$ is the electrical conductivity,
$\rho$ the mass density,
$\alpha$ the thermal expansion coefficient,
$g$ the acceleration due to gravity,
$\nu$ the kinematic viscosity, and
$\kappa$ the thermal diffusivity.
We apply no-slip BCs for the velocity at all boundaries, $\bm{u}=0$, constant temperatures at the end-faces, i.e., $T=T_+$ at the bottom plate at $z=0$ and $T=T_-$ at the top plate at $z=H$, and adiabatic BC at the side walls, $\partial T/\partial \bm{n}=0$, where $\bm{n}$ is the vector orthogonal to the surface.  
All solid boundaries are considered electrically insulated; Neumann BCs for the electric potential are $\partial \phi/\partial \bm{n}=0$.
The simulation domain is cubic of height $H$, width $W$, length $L$, $H=W=L$, i.e. has aspect ratio $\Gamma\equiv L/H = 1$. 
Most simulations are carried out for liquid metals such as {\em GaInSn} at $\Pran=0.025$, for $\Ra$ up to $10^9$ and $\Ha$ up to $2000$. Some DNS are conducted also for $\Pran=8$, to extend the parameter range studied in Ref.~\cite{Lim2019}.
In the bulk, spatial flow fluctuations are resolved down to 2-5, occasionally 10 Kolmogorov miscroscales, near the rigid walls we resolve the thermal and Hartmann BLs~\cite{Shishkina2010}. 
Our DNSs have been carried out with an MC extension of {\sc goldfish} \cite{Kooij2018,Reiter2022, Reiter2021a}.
Further details of the simulations are provided in appendix \ref{appendix}.

In Fig.~\ref{Fig3} we present $\Nu-1$ as a function of $\Ra$ (Fig.~\ref{Fig3}~a-b) and $\Ha$ (Fig.~\ref{Fig3}~c-d), respectively, from
our DNS, experimental \cite{Cioni2000, King2015, Zuerner2020, Xu2023} and DNS data \cite{Liu2018, Akhmedagaev2020, Xu2023} for liquid metals, $0.025\leq\Pr\leq0.029$ (Fig.~\ref{Fig3}~b,~d) and a fluid with $\Pran=8$ (Fig.~\ref{Fig3}~a,~c). 
In addition, in Fig.~\ref{Fig3}~(b,~d), we plot for comparison the DNS data \cite{Yan2019} for free-slip BCs.
In Fig.~\ref{Fig3}~(a-b) one can see that $\Nu$ generally increases with growing $\Ra$, but with different slopes for different $\Ha$, which are steeper for larger $\Ha$.  
In the double logarithmic plots of Fig.~\ref{Fig3}~(a-b), the curves of the $(\Nu-1)$-vs.-$\Ra$ dependences for different $\Ha$ approach
each other when $\Ra$ increases.
In Fig.~\ref{Fig3}~(c-d) one can see that $\Nu$ remains to be almost unaffected by the magnetic field for relatively small $\Ha$, but for a strong Lorentz force (large $\Ha$) they gradually decrease with growing $\Ha$.  
Here, again, the decreasing slopes are different for different $\Ra$, and the transition to the regime, where the heat transport is affected by the magnetic field, depends on $\Ra$.
In summary, the data in Fig.~\ref{Fig3} look rather different in different experiments and DNS. 
In what follows, we show that our model results in a collapse of all data points on a single master curve.

\begin{figure}\centering
\includegraphics[width = 16cm]{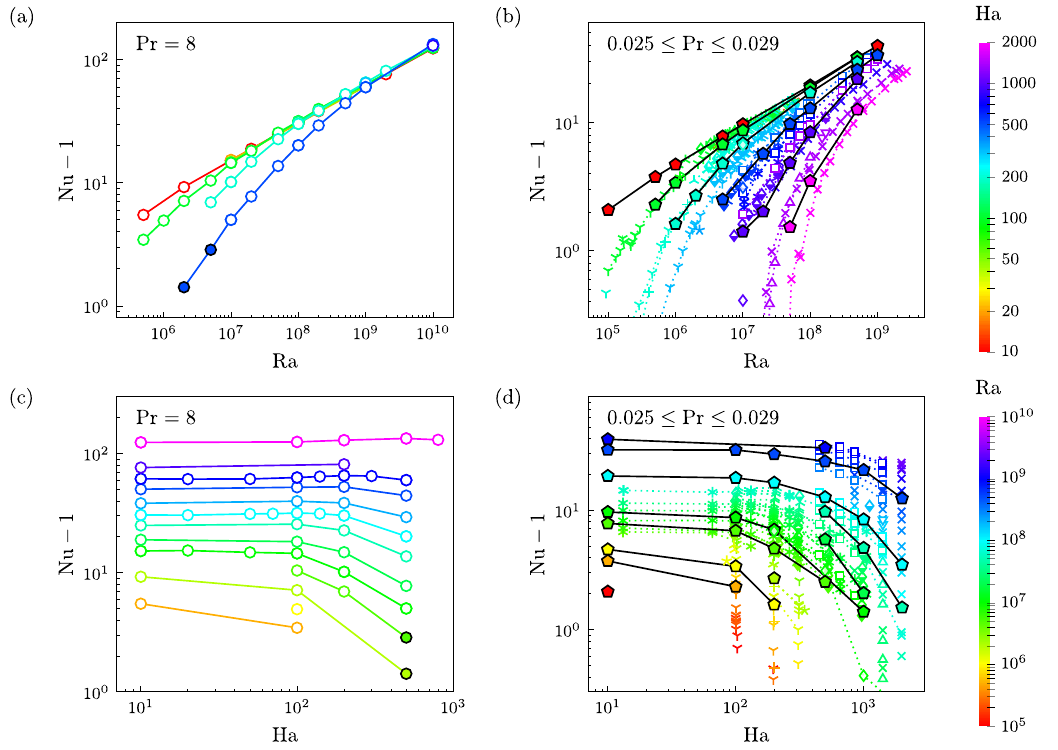}
\vskip2mm
\includegraphics[width = 16cm]{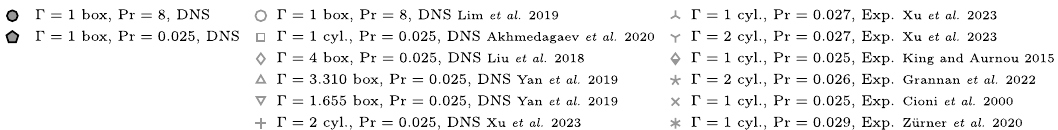}
\caption{The dimensionless convective heat transport, i.e., $\Nu-1$, as functions of (a, b) $\Ra$ and (c-d) $\Ha$, for (a, c) $\Pran=8$ and (b, d) $0.025\leq\Pr\leq0.029$.
The color scales are according to (a-b) $\Ha$ and (c-d) $\Ra$.}
\label{Fig3}
\end{figure}

\section{{Model validation}}
\begin{figure}\centering
\includegraphics[width = 16cm]{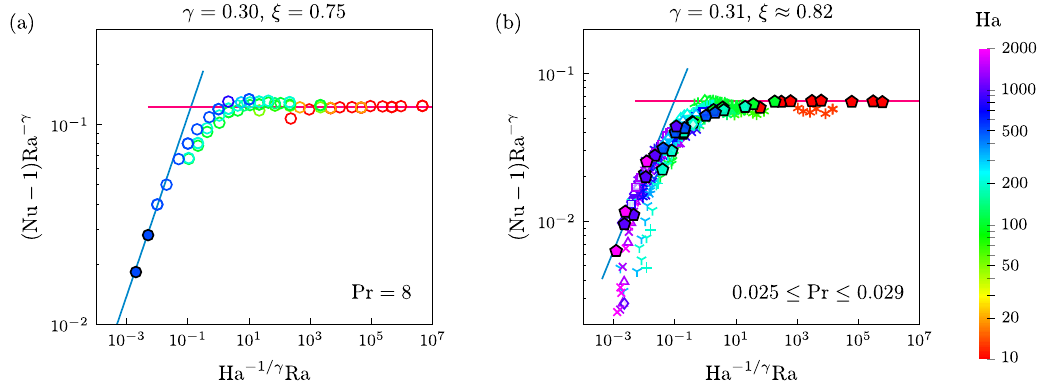}
\caption{All data from Fig.~\ref{Fig3} follow master scaling curves if plotted as Fig.~\ref{Fig2} suggests, for (a) $\Pran=8$ and (b)
$0.025\leq\Pran\leq0.029$. The values of $\gamma$ are calculated from GL theory, and the values of $\xi$ are calculated from Eq.~(\ref{qqq5}).  
Pink and blue lines show the predictions of the slopes in the buoyancy and Lorentz-force-dominated regimes, respectively.  
The symbols have the same meaning as in Fig.~\ref{Fig3}.}
\label{Fig4}
\end{figure}

We now validate the model using the data presented in Fig.~\ref{Fig3}.  
To calculate the scaling exponent $\gamma$ in the buoyancy-dominated regime, for the considered $\Pran$ and $\Ra$-ranges, we use GL theory, which gives $\gamma$ about 0.30 for $\Pran=8$ and about 0.31 for $\Pran=0.025$. 
These values agree very well with fits to data for the $\Ra$-ranges in the buoyancy-dominated regime for both values of $\Pran$, see appendix \ref{appendix} for further details. 
Using Eq.~(\ref{qqq5}) with $\gamma=0.30$ for $\Pran=8$ and $\gamma=0.31$ for $\Pran=0.025$, we calculate the exponent $\xi$ in the Lorentz-force-dominated regime, which equals $\xi=0.75$ for $\Pran=8$ and $\xi\approx0.82$ for $\Pran=0.025$.  
In Fig.~\ref{Fig4} we plot all data presented previously Fig.~\ref{Fig3} using the coordinates suggested by our model and visualised in Fig.~\ref{Fig2}.  
This results in a clear collapse of the data onto master curves for $\Pran=8$ (Fig.~\ref{Fig4}~a) and $0.025\leq\Pran\leq0.029$ (Fig.~\ref{Fig4}~b).  
Some deviation of the data from the master curve in Fig.~\ref{Fig4}~(b) is observed when $\Ra$ is relatively small and the flow is in the wall-mode regime, that is, before the onset of bulk convection where heat transport is confined to the near-wall region. 
Since the influence of no-slip side-walls is outside of the scope of our model, the observed deviations are expected in this regime.

\section{{Conclusion}}
In conclusion, we have proposed a heat transport model for the transition between the buoyancy- and Lorentz-force-dominated regimes of vertical MC.
We validated the model using our DNS and data available in the literature.
We wish to emphasize that the proposed model is parameter-free. 
For a given $\Pran$ and $\Ra$-range, one can calculate the scaling exponent in the buoyancy-dominated regime, using GL theory.
Then, using Eq.~(\ref{qqq5}), one can calculate the scaling exponent $\xi$ in the Lorentz-force-dominated regime and collapse the data on a master curve by rescaling the coordinate axes as in Fig.~\ref{Fig2}.
The model can in principle be extended to include the effect of a fluctuating magnetic field, this merely results in an adjustment of the prefactors. 

\section*{{Acknowledgements}}
The authors thank R.~E.~Ecke, D.~Lohse and G.~Vasil for fruitful discussions and R. Akhmedagaev for providing data and acknowledge the financial support from the Deutsche Forschungsgemeinschaft (SPP1881 "Turbulent Superstructures" and grants  Sh405/7, Sh405/16 and Li3694/1). 
This work used the ARCHER2 UK National Supercomputing Service (https://www.archer2.ac.uk).
Computing resources have been provided by the UK Turbulence Consortium (EPSRC grants EP/R029326/1 and EP/X035484/1).

\bibliographystyle{plain}
\bibliography{References}

\begin{thebibliography}{10}

\bibitem{Ahlers2022}
G.~Ahlers, E.~Bodenschatz, R.~Hartmann, X.~He, D.~Lohse, P.~Reiter, R.J.A.M.
  Stevens, R.~Verzicco, M.~Wedi, S.~Weiss, X.~Zhang, L.~Zwirner, and
  O.~Shishkina.
\newblock {Aspect ratio dependence of heat transfer in a cylindrical
  Rayleigh--B\'enard cell}.
\newblock {\em Phys. Rev. Lett.}, 128:084501, 2022.

\bibitem{Ahlers2009}
G.~Ahlers, S.~Grossmann, and D.~Lohse.
\newblock {Heat transfer and large scale dynamics in turbulent
  Rayleigh--B\'enard convection}.
\newblock {\em Rev. Mod. Phys.}, 81:503--537, 2009.

\bibitem{Akhmedagaev2020}
R.~Akhmedagaev, O.~Zikanov, D.~Krasnov, and J.~Schumacher.
\newblock {Turbulent Rayleigh--B\'enard convection in a strong vertical
  magnetic field}.
\newblock {\em J. Fluid Mech.}, 895:R4, 2020.

\bibitem{Chandrasekhar1961}
S.~Chandrasekhar.
\newblock {\em {Hydrodynamic and Hydromagnetic Stability}}.
\newblock Clarendon Press, 1961.

\bibitem{Cioni2000}
S.~Cioni, S.~Chaumat, and J.~Sommeria.
\newblock {Effect of a vertical magnetic field on turbulent Rayleigh--B\'enard
  convection}.
\newblock {\em Phys. Rev. E}, 62:R4520--R4523, 2000.

\bibitem{Davidson1999}
P.~A. Davidson.
\newblock {Magnetohydrodynamics in materials processing}.
\newblock {\em Annu. Rev. Fluid Mech.}, 31:273--300, 1999.

\bibitem{Davidson2016}
P.~A. Davidson.
\newblock {\em Introduction to Magnetohydrodynamics}.
\newblock Cambridge University Press, 2016.

\bibitem{Ecke2023}
R.~E. Ecke and O.~Shishkina.
\newblock {Turbulent rotating Rayleigh--B\'enard convection}.
\newblock {\em Annu. Rev. Fluid Mech.}, 55:603--638, 2023.

\bibitem{Grossmann2000}
S.~Grossmann and D.~Lohse.
\newblock {Scaling in thermal convection: A unifying theory}.
\newblock {\em J. Fluid Mech.}, 407:27--56, 2000.

\bibitem{Grossmann2001}
S.~Grossmann and D.~Lohse.
\newblock {Thermal convection for large Prandtl numbers}.
\newblock {\em Phys. Rev. Lett.}, 86:3316--3319, 2001.

\bibitem{Hartmann1937b}
J.~Hartmann and F.~Lazarus.
\newblock Hg-dynamics ii. experimental investigations on the flow of mercury in
  a homogeneous magnetic field.
\newblock {\em Det Kgl Danske Videnskabernes Selskkab Math-fys Medd},
  15(7):1--45, 1937.

\bibitem{Jones2011}
C.~A. Jones.
\newblock {Planetary magnetic fields and fluid dynamos}.
\newblock {\em Annu. Rev. Fluid Mech.}, 43:583--614, 2011.

\bibitem{King2015}
E.~M. King and J.~M. Aurnou.
\newblock {Magnetostrophic balance as the optimal state for turbulent
  magnetoconvection}.
\newblock {\em Proc. Natl. Acad. Sci.}, 112:990--994, 2015.

\bibitem{Kooij2018}
G.~L. Kooij, M.~A. Botchev, E.~M.A. Frederix, B.~J. Geurts, S.~Horn, D.~Lohse,
  E.~P. van~der Poel, O.~Shishkina, R.~J. A.~M. Stevens, and R.~Verzicco.
\newblock {Comparison of computational codes for direct numerical simulations
  of turbulent Rayleigh--B\'enard convection}.
\newblock {\em Comp. Fluids}, 166:1--8, 2018.

\bibitem{Lim2019}
Z.~L. Lim, K.~L. Chong, G.-Y. Ding, and K.-Q. Xia.
\newblock Quasistatic magnetoconvection: heat transport enhancement and
  boundary layer crossing.
\newblock {\em J. Fluid Mech.}, 870:519--542, 2019.

\bibitem{Liu2018}
W.~Liu, D.~Krasnov, and J.~Schumacher.
\newblock {Wall modes in magnetoconvection at high Hartmann numbers}.
\newblock {\em J. Fluid Mech.}, 849:R2, 2018.

\bibitem{Ni2012}
M.-J. Ni and J.-F. Li.
\newblock {A consistent and conservative scheme for incompressible MHD flows at
  a low magnetic Reynolds number. Part III: On a staggered mesh}.
\newblock {\em J. Comput. Phys.}, 231:281--298, 2012.

\bibitem{Prandtl1905}
L.~Prandtl.
\newblock {\"Uber Fl\"ussigkeitsbewegung bei sehr kleiner Reibung}.
\newblock In {\em Verhandlungen des III. Int. Math. Kongr., Heidelberg, 1904},
  pages 484--491. Teubner, 1905.

\bibitem{Reiter2021a}
P.~Reiter, O.~Shishkina, D.~Lohse, and D.~Krug.
\newblock {Crossover of the relative heat transport contributions of plume
  ejecting and impacting zones in turbulent Rayleigh--B\'nard convection}.
\newblock {\em Europhys. Lett.}, 134:34002, 2021.

\bibitem{Reiter2022}
P.~Reiter, X.~Zhang, and O.~Shishkina.
\newblock {Flow states and heat transport in Rayleigh--B\'enard convection with
  different sidewall boundary conditions}.
\newblock {\em J. Fluid Mech.}, 936:A32, 2022.

\bibitem{Schlichting1979}
H.~Schlichting.
\newblock {\em Boundary layer theory}.
\newblock McGraw-Hill, 1979.

\bibitem{Shishkina2021}
O.~Shishkina.
\newblock {Rayleigh--B\'enard convection: The container shape matters}.
\newblock {\em Phys. Rev. Fluids}, 6:090502, 2021.

\bibitem{Shishkina2010}
O.~Shishkina, R.~J. A.~M. Stevens, S.~Grossmann, and D.~Lohse.
\newblock Boundary layer structure in turbulent thermal convection and its
  consequences for the required numerical resolution.
\newblock {\em New J. Phys.}, 12:075022, 2010.

\bibitem{Stevens2013}
R.~J. A.~M. Stevens, E.~P. van~der Poel, S.~Grossmann, and D.~Lohse.
\newblock {The unifying theory of scaling in thermal convection: The updated
  prefactors}.
\newblock {\em J. Fluid Mech.}, 730:295--308, 2013.

\bibitem{Weiss2014}
N.~O. Weiss and M.~R.~E. Proctor.
\newblock {\em {Magnetoconvection}}.
\newblock Cambridge University Press, 2014.

\bibitem{Xu2023}
Y.~Xu, S.~Horn, and J.~M. Aurnou.
\newblock The transition from wall modes to multimodality in liquid gallium
  magnetoconvection.
\newblock {\em https://arxiv.org/pdf/2303.08966.pdf}, 2023.

\bibitem{Yan2019}
M.~Yan, M.~A. Calkins, S.~Maffei, K.~Julien, S.~Tobias, and P.~Marti.
\newblock Heat transfer and flow regimes in quasi-static magnetoconvection with
  a vertical magnetic field.
\newblock {\em J. Fluid Mech.}, 877:1186--1206, 2019.

\bibitem{Zuerner2020}
T.~Z\"urner, F.~Schindler, T.~Vogt, S.~Eckert, and J.~Schumacher.
\newblock Flow regimes of rayleigh--b\'enard convection in a vertical magnetic
  field.
\newblock {\em J. Fluid Mech.}, 894:A21, 2020.

\end{thebibliography}

\appendix
\section{Appendix} \label{appendix}

\subsection{Numerical simulations}
All direct numerical simulations have been carried using the direct numerical solver {\sc goldfish} \cite{Kooij2018}, which has been widely used in previous studies of different convective flows.
The new version of the code that applies a fourth-order finite-volume discretisation on staggered grids and a third-order Runge--Kutta time marching scheme \cite{Reiter2022, Reiter2021a} has been extended to simulate magnetoconvective flows, where a consistent and conservative scheme~\cite{Ni2012} is utilised to calculate the current density and the Lorentz force.
The three-dimensional DNS are performed using staggered grids that provide fine resolution in the core part of the domain and near the rigid walls~\cite{Shishkina2010}, to resolve the thermal and Hartmann boundary layers.
The main parameters and key observables of the simulations are summarised in table~\ref{tbl:simulations}.

The DNS dataset comprises 37 simulations to cover the necessary ranges in $\Ha$ and $\Ra$. To obtain a dataset this large within the available resources, a compromise had to be made in terms of the bulk resolution. Hence a number of simulations do not resolve scales close to the Kolmogorov microscale. 
To ensure accurate predictions of mean $\Nu$, grid refinement studies have been done for key simulations (marked by * in the $h_{\text{DNS}}/h_{\text{K}}$ column).  For these cases, changing the grid resolution by a factor of at least 2.5 resulted in changes of $\Nu$ by less than 1\% over a long-time average.

\subsection{Grossmann-Lohse scaling} 
\begin{figure}[h]
\centering
\includegraphics[height=4.75cm]{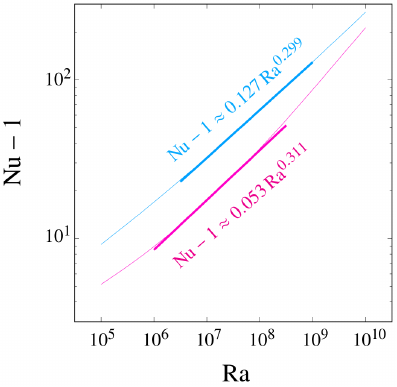}
\caption{Predictions of GL theory \cite{Grossmann2000, Grossmann2001, Stevens2013} (thin lines) for $\Nu-1$ versus $\Ra$ scaling relations in RBC without magnetic field, for $\Pr=8$ (blue) and $\Pr=0.025$ (pink).
The power fits for the considered $\Ra$-ranges are highlighted with thick lines of the corresponding colors.}
\label{fig:GL}
\end{figure}

Figure \ref{fig:GL} shows a comparison between the predictions of Grossmann-Lohse theory \cite{Grossmann2000, Grossmann2001, Stevens2013} for the dependence of the dimensionless heat transport $\Nu-1$ on the Rayleigh number $\Ra$, $\Nu-1 = A \, \Ra^\gamma$ for $\Pr$-dependent prefactor $A$. 
Fits to data have been carried out for 
$3\times 10^6 \leqslant 10^9$ for $\Pr=0.025$
and $3\times 10^6 \leqslant 10^9$ for $\Pr = 8$. As can been seen from the data presented in fig.~\ref{fig:GL}, the theoretical predictions agree very well with the data. 

\subsection{Flow and temperature field visualisations}
\begin{figure}[h]
    \centering
    \includegraphics[width = 15cm]{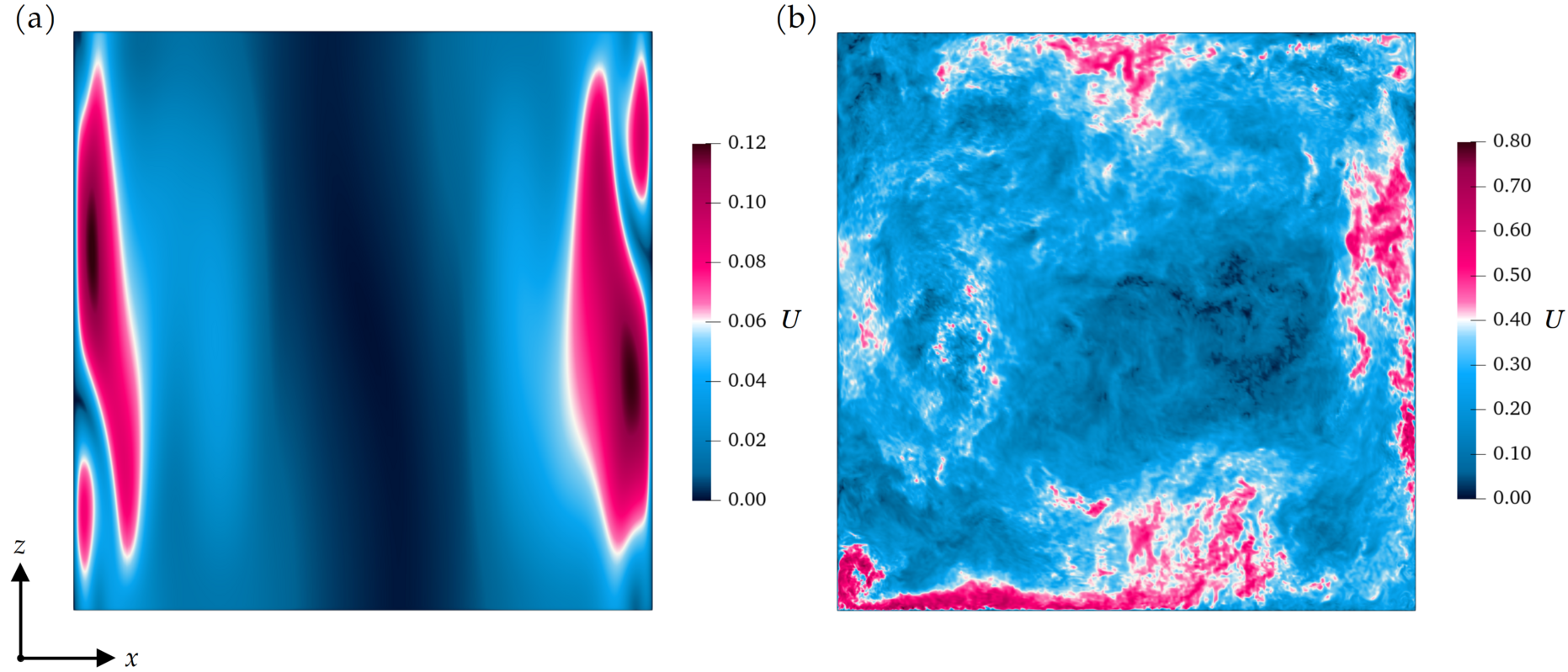}
    \caption{Instantaneous velocity magnitude $U$ shown on the $y$ mid-plane for (a) $Ra = 10^7$, $Ha = 1000$ and (b) $Ra = 10^9$, $Ha = 10$.}
    \label{fig:rbc-512-par-eff-u}
\end{figure}

\begin{figure}[h]
    \centering
    \includegraphics[width = 15cm]{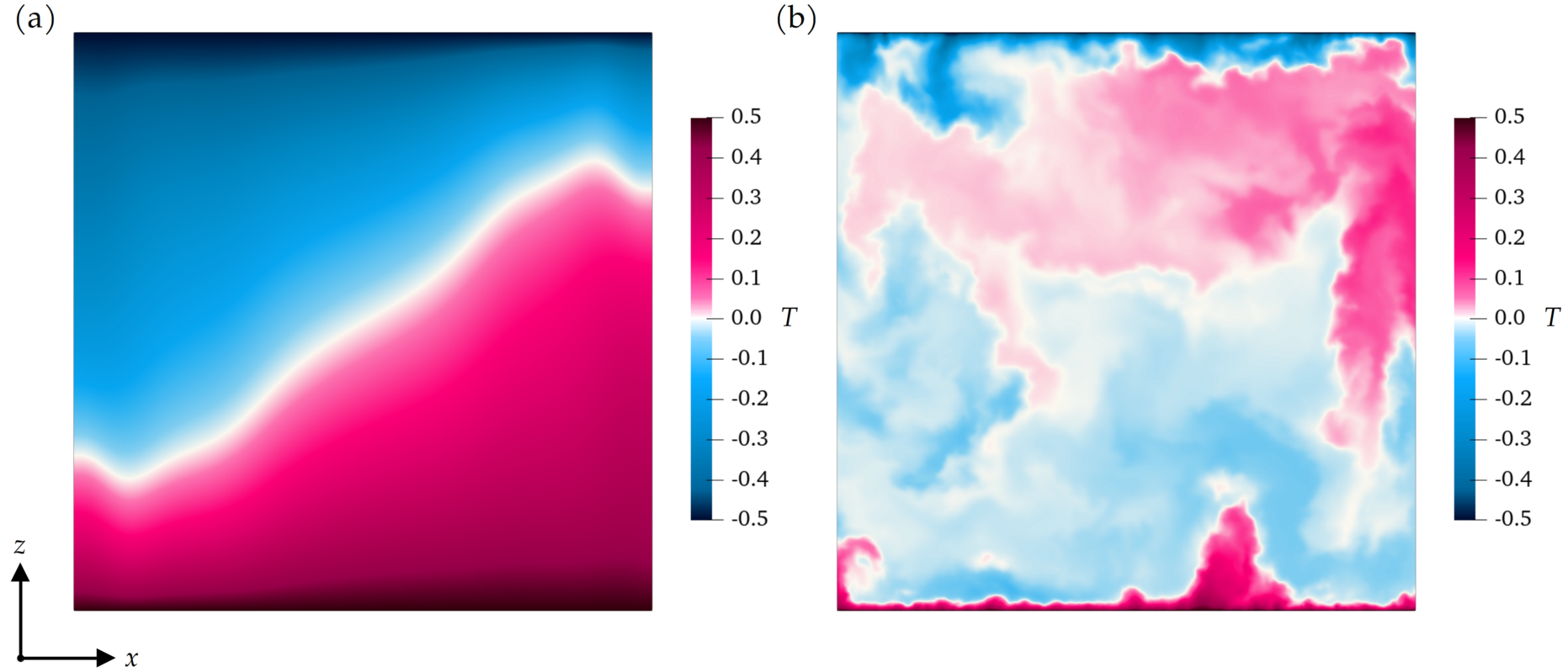}
    \caption{Instantaneous temperature field $T$ shown on the $y$ mid-plane for (a) $Ra = 10^7$, $Ha = 1000$ and (b) $Ra = 10^9$, $Ha = 10$.}
    \label{fig:rbc-512-par-eff-T}
\end{figure}

Figures \ref{fig:rbc-512-par-eff-u} and \ref{fig:rbc-512-par-eff-T} present visualisations the velocity magnitude and temperature, respectively, at an instant in time during statistically steady evolution in the magnetically dominated and the buoyancy-dominated regimes. In both figures, the respective panels (a) correspond to 
$\Ra = 10^7$, $\Ha = 1000$, panels (b) to $\Ra = 10^9$, $\Ha = 10$.
As can be seen from a comparison of the flow and temperature field between both regimes, both velocity and temperature fluctuate on much smaller scales in the buoyancy-dominated regime. The magnetically dominated regime is shows strong vertical flows near the walls, and we note the absence of plumes in the temperature field.  

\begin{table}
	\begin{center}
		\def~{\hphantom{0}}
		\small
		\begin{tabular}{cccccccccccccc}

            $\Pr$ &$\Ha$ & $\Ra$ & $\Nu$ & $\sigma_{\Nu}$ & $N_x$ & $N_y$ & $N_z$& $T_{run}$ & 
            $\mathcal{N}_\theta$ &$\mathcal{N}_{\text{Ha}}$ & 
             $h_{\text{DNS}}/h_{\text{K}}$\\ \hline
                
		$8$ & 500 & $2.0 \times 10^6$ & 2.42 & 0.01 & 100  & 100  & 300 & 350 & 90 & 8 & 0.40  \\ 
		        & 500 & $5.0 \times 10^6$ & 3.86 & 0.02 & 100  & 100  & 300 & 550 & 70 & 8 & 0.60\\ \hline

            0.025    & 10   & 1.0$\times10^5$ & 3.08  & 0.04 & 150 & 150 & 200 & 40  & 49 & 36 & 1.06 \\
                     & 10   & 5.0$\times10^5$ & 4.76  & 0.21 & 150 & 150 & 200 & 45  & 37 & 36 & 1.76 \\
                     & 10   & 1.0$\times10^6$ & 5.70  & 0.32 & 150 & 150 & 200 & 65  & 33 & 36 & 2.20 \\
                     & 10   & 5.0$\times10^6$ & 8.77  & 0.67 & 150 & 150 & 200 & 100 & 25 & 36 & 3.64 \\
                     & 10   & 1.0$\times10^7$ & 10.68 & 0.78 & 150 & 150 & 200 & 160 & 22 & 36 & 4.49* \\
                     & 10   & 1.0$\times10^8$ & 20.42 & 1.98 & 250 & 250 & 350 & 50  & 29 & 67 & 5.42 \\
                     & 10   & 5.0$\times10^8$ & 33.18 & 1.47 & 250 & 250 & 350 & 90  & 21 & 67 & 8.76 \\
                     & 10   & 1.0$\times10^9$ & 40.47 & 2.32 & 350 & 350 & 450 & 40  & 25 & 87 & 8.64* \\
                     & 100  & 5.0$\times10^5$ & 3.29  & 0.06 & 200 & 200 & 250 & 20  & 60 & 10 & 1.07 \\
                     & 100  & 1.0$\times10^6$ & 4.39  & 0.14 & 200 & 200 & 250 & 30  & 50 & 10 & 1.34 \\
                     & 100  & 5.0$\times10^6$ & 7.74  & 0.56 & 200 & 200 & 250 & 40  & 35 & 10 & 2.40 \\
                     & 100  & 1.0$\times10^7$ & 9.69  & 0.44 & 200 & 200 & 250 & 60  & 31 & 10 & 3.09* \\
                     & 100  & 1.0$\times10^8$ & 19.72 & 1.36 & 220 & 220 & 300 & 45  & 24 & 13 & 5.81 \\
                     & 100  & 5.0$\times10^8$ & 33.10 & 1.76 & 250 & 250 & 350 & 65  & 21 & 16 & 8.45* \\
                     & 200  & 1.0$\times10^6$ & 2.62  & 0.03 & 220 & 220 & 350 & 120 & 101 & 15 & 1.13 \\
                     & 200  & 2.0$\times10^6$ & 3.69  & 0.01 & 220 & 220 & 350 & 225 & 84 & 15 & 1.46 \\
                     & 200  & 5.0$\times10^6$ & 5.78  & 0.22 & 220 & 220 & 350 & 225 & 66 & 15 & 2.06 \\
                     & 200  & 1.0$\times10^7$ & 7.87  & 0.39 & 220 & 220 & 350 & 160 & 56 & 15 & 2.65* \\
                     & 200  & 1.0$\times10^8$ & 18.06 & 1.82 & 250 & 250 & 400 & 100 & 42 & 17 & 5.07 \\
                     & 200  & 5.0$\times10^8$ & 30.56 & 2.63 & 250 & 250 & 400 & 100 & 32 & 17 & 8.66* \\
                     & 500  & 5.0$\times10^6$ & 3.51  & 0.06 & 220 & 220 & 350 & 650 & 86 & 9 & 1.82 \\
                     & 500  & 2.0$\times10^7$ & 6.70  & 0.34 & 220 & 220 & 350 & 190 & 61 & 9 & 3.02 \\
                     & 500  & 5.0$\times10^7$ & 10.73 & 0.64 & 220 & 220 & 350 & 300 & 48 & 9 & 4.27* \\
                     & 500  & 1.0$\times10^8$ & 13.88 & 0.76 & 250 & 250 & 400 & 500 & 48 & 11 & 4.74 \\
                     & 500  & 5.0$\times10^8$ & 26.75 & 1.77 & 250 & 250 & 400 & 150 & 34 & 11 & 8.36 \\
                     & 500  & 1.0$\times10^9$ & 34.44 & 2.59 & 250 & 250 & 400 & 250 & 30 & 11 & 10.64* \\
                     & 1000 & 1.0$\times10^7$ & 2.41  & 0.26 & 220 & 220 & 300 & 600 & 90 & 5 & 1.92* \\
                     & 1000 & 2.0$\times10^7$ & 3.02  & 0.08 & 220 & 220 & 300 & 1300 & 80 & 5 & 2.30 \\
                     & 1000 & 5.0$\times10^7$ & 5.84  & 0.37 & 220 & 220 & 300 & 500 & 56 & 5 & 3.02 \\
                     & 1000 & 1.0$\times10^8$ & 9.39  & 0.51 & 220 & 220 & 300 & 1100 & 44 & 5 & 3.77 \\
                     & 1000 & 5.0$\times10^8$ & 22.78 & 1.35 & 220 & 220 & 300 & 60  & 18 & 5 & 6.44* \\
                     & 2000 & 5.0$\times10^7$ & 2.53  & 0.09 & 250 & 250 & 400 & 90  & 116 & 5 & 2.20 \\
                     & 2000 & 1.0$\times10^8$ & 4.50  & 0.11 & 250 & 250 & 400 & 160 & 86 & 5 & 2.68 \\
                     & 2000 & 5.0$\times10^8$ & 13.60 & 0.63 & 250 & 250 & 400 & 220 & 49 & 5 & 4.47* \\
            
		\end{tabular}
		\caption{Details on the conducted DNS, including the 
        standard deviation $\sigma_{\small \Nu}$ of the Nusselt number $\Nu$, the
        number of nodes $N_x$, $N_y$, $N_z$ in the directions $x$, $y$ and $z$, respectively; the number of simulation free-fall times used for averaging $T_{run}$;
		the number of the nodes within the thermal boundary layer, 
		and within the Hartmann boundary layer $\mathcal{N}_{\text{Ha}}$;
		the Kolmogorov microscale, $h_{\text{K}}$,
		and the relative mean grid stepping, $h_{\text{DNS}}/h_{\text{K}}$. 
            Simulations marked by * are those for which grid refinement studies have been completed.
		}
		\label{tbl:simulations}
	\end{center}
\end{table}

\end{document}